\documentclass[preprint2]{aastex}

\usepackage{amsmath}
\usepackage{amsfonts}
\usepackage{amssymb}
\usepackage{graphicx}
\usepackage{xcolor}
\usepackage{epstopdf}
\usepackage[colorlinks=true,citecolor=black,linkcolor=black,urlcolor=blue,citebordercolor=white,linkbordercolor=white,urlbordercolor=white]{hyperref}
\usepackage{natbib}
\usepackage{multirow}

\begin{document}

\title{MMTF Discovery of Giant Ionization Cones in MR 2251$-$178:\\
  Implications for Quasar Radiative Feedback}

\author{Kory Kreimeyer\altaffilmark{1} \& Sylvain
  Veilleux\altaffilmark{1,2,3}} 

\altaffiltext{1}{Department of Astronomy, University of Maryland,
  College Park, MD 20742, USA; kory@astro.umd.edu,
  veilleux@astro.umd.edu}

\altaffiltext{2}{Astroparticle Physics Laboratory, NASA Goddard Space
  Flight Center, Greenbelt, MD 20771, USA}

\altaffiltext{3}{Max-Planck-Institut f\"ur extraterrestrische Physik,
  Postfach 1312, D-85741 Garching, Germany}


\begin{abstract}
  We report the discovery of giant ionization cones in the 140-kpc
  nebula around quasar MR~2251$-$178 based on deep [\ion{O}{3}]
  $\lambda$5007/H$\beta$ and [\ion{N}{2}] $\lambda$6583/H$\alpha$ flux
  ratio maps obtained with the Maryland-Magellan Tunable Filter (MMTF)
  on the Baade-Magellan 6.5m telescope. These cones are aligned with
  the weak double-lobed radio source observed on smaller scale ($<$30
  kpc). They have an opening angle $\sim$120$\degr\pm$10$\degr$ and
  subtend $\sim$65--90\% of $4\pi$ steradians, where the large
  uncertainty takes into account possible projection effects.  The
  material in the outer ionization cones is matter-bounded, indicating
  that all ionizing photons emitted through the cones escape from the
  system. The quasar ionizing flux is $\sim$2-3 times fainter outside
  of these cones, despite the largely symmetric geometry of the nebula
  in [\ion{O}{3}].  Overall, adding up the contributions from both
  inside and outside the cones, we find that $\sim$65--95\% of the
  quasar ionizing radiation makes its way out of the system. These
  results emphasize the need for line ratio maps to quantify the
  escape fraction of ionizing radiation from quasars and the
  importance of quasar radiative feedback on the intergalactic medium.
\end{abstract}
\maketitle

\section{Introduction}

There is growing evidence that feedback from active galactic nuclei
(AGN) plays a significant role in producing the galaxy populations we
observe today \citep[e.g.][and references
therein]{2005ARA&A..43..769V,2012ARA&A..50..455F}. AGN feedback also
seems to be needed to explain the scaling relations between black hole
and spheroid masses
\citep[e.g.][]{2000ApJ...539L...9F,2000ApJ...539L..13G}. This feedback
may take two forms: mechanical (winds or jets) or radiative. While
observational evidence for powerful AGN-driven outflows has grown in
recent years \citep[e.g.][and references
therein]{2013ApJ...768...75R}, there is no doubt that the ionizing
radiation from the AGN also alters the gas in the host and sometimes
much beyond
\citep[e.g.,][]{1988ApJ...327..570B,2003AJ....126.2185V,2008ApJ...676..816G,2012ApJ...759..117C}.
Quasar radiative feedback is likely responsible for He reionization at
$z \sim 3$ \citep[e.g.,][and references therein]{2004ApJ...605..631Z}
and at least partly responsible for H reionization at $z \ga$~6
\citep[e.g.,][]{2001AJ....122.2850B,2001ARA&A..39...19L,2002AJ....123.1247F,2011ApJS..192...18K}. It
may also be affecting star formation in galaxies near the quasars
\citep[e.g.,][]{2004MNRAS.353..301F,2012MNRAS.421.2543B,2012MNRAS.422.2980S,2012ApJ...756..150D}.

The fraction of ionization radiation that escapes from the central
engine and deposits its energy in the host or intergalactic
medium (IGM) depends on the exact topology of the host ISM and thus
likely varies significantly from one object to the next. The detection
of Lyman continuum emission in some quasars does not constrain the
overall (4 $\pi$ steradians) escape fraction since it does not take
into account likely angular dependences. The ionization structure of
the host ISM is in principle a good probe of the geometry of the
ionizing radiation field, but the ISM is often also affected by
mechanical feedback
\citep[e.g][]{2011MNRAS.418.2032V,2011ApJ...732....9G}, making the
interpretation more difficult. An important exception is the giant
(140 kpc\footnotemark[1]) photoionized nebula around the type 1
radio-quiet quasar MR~2251$-$178
\citep{1990ApJ...356..389M,1999ApJ...524L..83S}. To our knowledge this
is the largest emission-line nebula around any radio-quiet AGN
\citep[e.g.,][]{2010A&A...519A.115H}.  This nebula is in rotation
around the host, unaffected dynamically by the weak, double-lobed
radio source observed to extend $\sim$27 kpc centered on the nucleus
\citep{1983MNRAS.202..125B,1990ApJ...356..389M,1999ApJ...524L..83S}.
The central quasar ($L_{\rm UV} \sim 7 \times 10^{21}$ W Hz$^{-1}$,
corrected for Galactic extinction) seems solely responsible for the
high ionization level of the nebula. At a distance of only 263
Mpc\footnotemark[1], corresponding to a scale of $\sim$1.3 kpc
arcsec$^{-1}$, this nebula is thus an excellent laboratory to study
the effects of radiative feedback.  \footnotetext[1]{Based on a
  redshift $z$ = 0.0640 and a cosmology with $H_0$ = 73 km s$^{-1}$
  Mpc$^{-1}$, $\Omega_{\rm matter}$ = 0.27, and $\Omega_{\rm vacuum}$
  = 0.73.  The size of the nebula reported here is consistent with
  previous studies that reported nebular sizes $\sim$200 kpc based on
  $H_0=50$ km s$^{-1}$ Mpc$^{-1}$.}

In this {\em Letter}, we map for the first time the ionization
structure of the nebula around MR~2251$-$178 using deep narrowband
images centered on key optical emission lines. We describe the
observations and methods used to reduce the data in \S 2. The images
of MR~2251$-$178 are presented in \S 3 along with line ratio maps
created from these images. In \S 4, these line ratio maps are compared
with photoionization models to derive the ionization structure of the
nebula around MR~2251$-$178 and the fraction of ionizing radiation
that escapes from the system as a function of azimuth angle.  These
results are discussed in the context of quasar radiative feedback.

\section{Observations and Data Reduction}

Data on MR~2251$-$178 were obtained in May and August 2011 with the
Maryland-Magellan Tunable Filter \citep[MMTF,][]{2010AJ....139..145V}
in the Inamori-Magellan Areal Camera \& Spectrograph
\citep[IMACS,][]{2011PASP..123..288D} on the 6.5~m Magellan-Baade
Telescope.  MMTF allows narrowband (FWHM$\sim$15 \AA) imaging at a
wavelength between $\sim 5000-9200$ \AA\ over a large field of view
($\sim$27$\times$27$^\prime$, monochromatic within $\sim$5$^\prime$ of
the optical axis).  Multiple 20-minute exposures centered on
redshifted [\ion{O}{3}] $\lambda$5007, H$\alpha$, and [\ion{N}{2}]
$\lambda$6583 were taken for a total integration time of 60, 100, and
140 minutes, respectively.  Additional images centered on 5044 \AA\
and 6621 \AA\ were obtained immediately before or after the line
images to map the line-free continuum emission.

All images were reduced using the standard MMTF pipeline
software\footnote{Available from
  \url{http://www.astro.umd.edu/~veilleux/mmtf/datared.html}.}.  
A photometric calibration was applied to each image, based on
observations of a photometric standard star taken on the same night.
The continuum images were scaled and subtracted from the line images
to obtain the pure emission-line maps.  All images were corrected for
foreground Galactic extinction, following the procedure of
\citet{1989ApJ...345..245C} with extinction parameter $R_V=3.07$, but
no attempt was made to correct for extinction within the host
galaxy. The images were smoothed using the \textit{adaptsmooth}
code\footnote{Available at
  \url{http://www.mpia.de/~zibetti/software/adaptsmooth.html}.}
described by \citet{2009arXiv0911.4956Z}.  This program replaces each
pixel of the image with the median value of all pixels within a
variable-sized smoothing radius, chosen as the minimum radius which
provides $S/N>3$.  The smoothing enhanced the visibility of the
diffuse gas emission while leaving the bright galaxy emission mostly
unaffected.

\section{Results: Giant Ionization Cones}

Figure~1 shows the adaptively-smoothed emission-line images of
MR~2251$-$178 as well as the unsmoothed narrowband continuum images.
The nebula extends out to $\sim$90 kpc from the quasar, elongated
along the east-west direction, lopsided to the south, and with a total
wing span of $\sim$140 kpc. Interestingly, these data do not reveal
any line emission beyond the extent of the nebula detected by
\citet{1999ApJ...524L..83S}, suggesting that we have detected the true
edge of the nebula. The superior image quality of the MMTF data over
those of Shopbell et al.\ resolves the nebula into a spectacular
complex of clumps and filaments down to scales of $\sim$1 kpc,
primarily distributed into a vast two-armed structure seemingly
detached from the central host.

\begin{figure}[t]
  \epsscale{0.8}
  \plotone{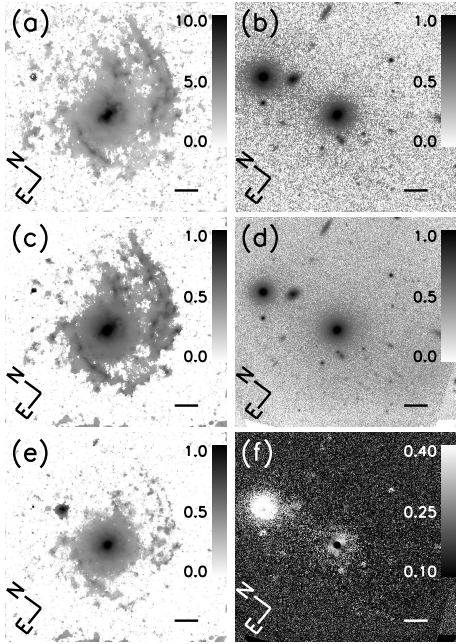}
  \caption{ MMTF images of MR~2251$-$178: (a) adaptively smoothed
    continuum-subtracted [\ion{O}{3}] $\lambda$5007 emission, (b)
    continuum emission at 5044 \AA, (c) adaptively smoothed
    continuum-subtracted H$\alpha$ emission, (d) continuum emission at
    6621 \AA, (e) adaptively smoothed continuum-subtracted
    [\ion{N}{2}] $\lambda$6583 emission, (f) Color map derived from
    the log ratio of (b) over (d). Residuals from the bright central
    source have been masked. Lighter shade indicates bluer color.  The
    bar at the bottom right corner of each panel indicates 20 kpc. The
    units in panels $a - e$ are $10^{-14}$ erg~s$^{-1}$ cm$^{-2}$
    arcsec$^{-2}$.  }
  \label{fig:fig1}
\end{figure}

The green (5044 \AA) and red (6621 \AA) continuum maps are shown in
Figures~1$b$ and 1$d$, while a color image derived from their ratios
is shown in Figure~1$f$. This last panel indicates a slight red excess
along the north-south direction, passing through the nucleus. While
stellar population may contribute to radial color gradients, we favor
reddening due to dust to explain this non-axisymmetric color
distribution.  At a distance of $\sim$5--10 kpc from the nucleus ({\em
  i.e.} well outside of possible artifacts due to the bright central
source), we measure a ratio
[log$(f_{5044}/f_{6621})_{\rm cone}$~/~log$(f_{5044}/f_{6621})_{\rm anti-cone}$] $\sim$
1.7, corresponding to $A_V=1.1\pm 0.5$
mag. \citep{1989ApJ...345..245C}.

We next constructed maps of MR~2251$-$178 in two of the classic
diagnostic line ratios: [\ion{O}{3}] $\lambda$5007/H$\beta$ and
[\ion{N}{2}] $\lambda$6583/H$\alpha$ (Figure~2), assuming that
$f($H$\beta) = f($H$\alpha) / 2.85$, the Case B recombination value
\citep{1989agna.book.....O}. We have neglected intrinsic dust
reddening because it is presumed to be small outside of the nucleus
based on our continuum color map (Fig. 1$f$) and the published
long-slit spectroscopy \citep[e.g.][]{1990ApJ...356..389M}. The line
ratio maps reveal a previously unknown ionization bicone with an
apparent opening angle $\sim$120$\degr\pm$10$\degr$ roughly aligned
along the east-west direction ($\sim$87$\degr$ west of north). The
biconical ionization structure is particularly evident within the
inner 30 kpc, but also extends all the way to the edge of the nebula
to encompass most of the gas in the large two-armed structure.
Interestingly, the double-lobed radio structure reported by
\citet{1990ApJ...356..389M} is aligned along the same direction
(within $\sim$10$\degr$) as the ionization cones and has a total
extent of $\sim$27 kpc \citep[][scaled to current
cosmology]{1990ApJ...356..389M} that is similar to that of the inner
cones (see white line in Figure 2). While this radio structure is not
believed to be a major source of kinetic energy to the nebula
\citep{1983MNRAS.202..125B}, it is clear that the ionizing radiation
from the quasar is preferentially escaping along the radio jet axis.

\begin{figure}[t]
  \epsscale{0.7}
  \plotone{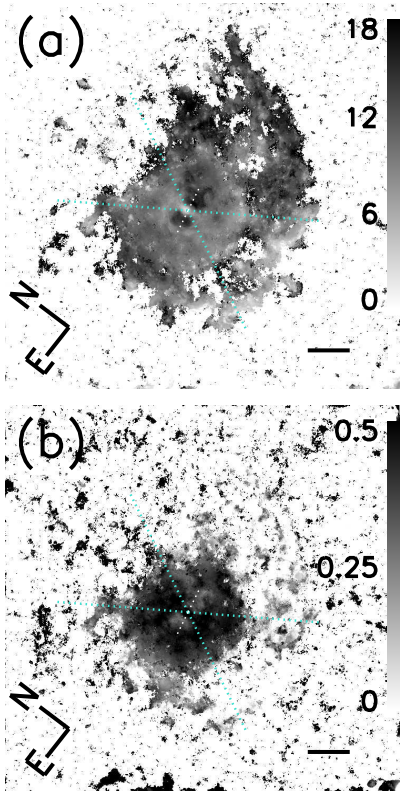}
  \caption{ Line ratio maps of MR~2251$-$178:
    (a) [\ion{O}{3}] $\lambda$5007/H$\beta$ and (b) [\ion{N}{2}]
    $\lambda$6583/H$\alpha$. We assumed $f(H\alpha)/f(H\beta)=2.85$
    (Case B recombination) to get $f(H\beta)$.  The bar at the bottom
    right corner of each panel indicates 20 kpc. The cyan dashed lines
    trace the ionization cones. The white dashed line indicates the
    orientation and extent of the radio structure reported by
    \citet{1990ApJ...356..389M}.  }
  \label{fig:fig2}
\end{figure}



\section{Discussion}

\subsection{Geometry of the Ionization Cones}

The inclination of the bicone axis with respect to our line of sight
is challenging to determine.  If it coincides with the radio jet axis
(\S 3), then the bisymmetry of the radio structure favors a large
inclination angle.  However, the Type~1 classification of this quasar
indicates that our line of sight avoids the central obscuring
torus. If the geometry of the ionization cones reflects the structure
of this inner torus, then the jet axis has to be $\gtrsim$30$\degr$
from the plane of the sky to allow our line of sight to lie within one
of the cones.

The three-dimensional distribution of the extended gas surrounding
MR~2251$-$178 is an unresolved topic with some favoring a spherical
envelope \citep[e.g.][]{1999ApJ...524L..83S}; however, the spherical
gas simulations of \citet{1996ApJ...467..197M}, when viewed from an
angle within the cones, have difficulties reproducing the V-shaped
biconical structure of MR~2251$-$178.  For this reason, we favor a
flattened distribution for the extended gas. Given the probability
distribution of apparent opening angles from a bicone with a true
opening angle of 70$\degr$ illuminating a thin disk \citep[Figure~8
of][]{1996ApJ...467..197M}, we estimate that the true opening angle of
the bicone in MR~2251$-$178 lies within the range
80$\degr-$130$\degr$.

\subsection{Line Ratio Diagrams}

The large [\ion{O}{3}]/H$\beta$ and small [\ion{N}{2}]/H$\alpha$
ratios throughout the nebula of MR~2251$-$178 suggest photoionization
by the quasar. This is confirmed in Figure~3, where these line ratios
are plotted in one of the diagnostic diagrams of
\citet{1981PASP...93....5B} and \citet{1987ApJS...63..295V}.
%
A more quantitative analysis of these data requires comparing the line
ratios with detailed AGN photoionization models. This is done in
Figure 3$c$--$d$, where we plot two photoionization models from
\citet{2004ApJS..153....9G}. The models assume a density
$n_H=1000~cm^{-3}$ and an incident ionizing power-law spectrum
($f_{\nu}\propto\nu^{\alpha_\nu}$) with index $\alpha_\nu=-1.4$ in the
range $5~eV\le h\nu\le1~keV$.  Figure~3$c$ shows the case for a dusty
ISM, while Figure~3$d$ presents the dust-free case.  The plot covers a
parameter space for the models in both metallicity and ionization
parameter
\begin{equation}
U\equiv\frac{\Phi}{n_ec}\label{eqn:eqn1}
\end{equation}
where $\Phi$ is the number of ionizing photons per unit area per unit
time, $n_e$ is the electron density at the front of the cloud, and $c$
is the speed of light.

\begin{figure}[t]
  \epsscale{1.}
  \plotone{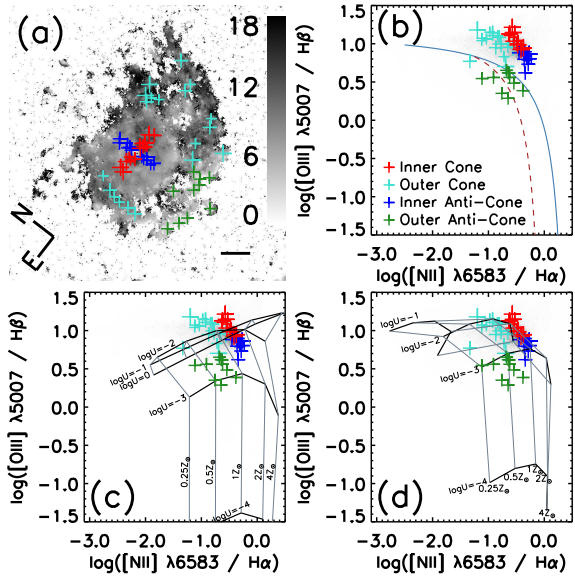}
  \caption{ (a) Same as Figure~2$a$, but with markers showing the
    locations of individually examined regions.  (b) The
    [\ion{O}{3}]/H$\beta$ vs [\ion{N}{2}] $\lambda$6583/H$\alpha$ line
    ratio diagram for the regions highlighted in (a).  The size of the
    crosses indicates their approximate uncertainty, with the largest
    symbols having an uncertainty of $\sim$10\% and the smallest
    symbols having an uncertainty of up to $\sim$ 30\%.  The solid
    blue line and brown dashed line are the starburst limits of
    \citet{2001ApJS..132...37K} and \citet{2003MNRAS.346.1055K},
    respectively.  The bottom two panels are the same diagnostic
    diagram with models from \citet{2004ApJS..153....9G} overlaid for
    the (c) dusty and (d) dust-free cases.  
  }
  \label{fig:fig3}
\end{figure}

Table 1 lists the range of parameters needed to explain the measured
line ratios.  In general, the dust-free models fit the line ratios
significantly better than the dusty models (see Figure~3), so we put
more weight on the results based on the dust-free models. In these
models, the ionization parameter is on average $\sim$0.3-0.5 dex
higher in the inner bicone than in the inner anti-cone region,
indicating that the gas along the east-west axis is exposed to an
ionizing flux that is $\sim$2-3 times stronger than along the
north-south axis (assuming the electron density is the same in the
cone and anti-cone regions).  The gas in the outer ``arms'' is
characterized by elevated ionization parameters, about 0.3 dex higher
than in the inner bicone, despite the unavoidable $\sim r^{-2}$
dilution of the quasar ionization radiation at these large radii. The
nebula seems matter-bounded with no evidence of ionization edges
(which would be characterized by a sudden drop in $U$).  There is also
a hint that the outer gas has slightly lower metallicity than the gas
in the inner region, suggesting either an external origin to this gas
or a smooth radial metallicity gradient \citep[consistent with the
most plausible origins for this gas: tidal debris from a recent galaxy
interaction and giant HI envelope ionized by, and gravitationally
bound to, the quasar. See][]{1999ApJ...524L..83S}.

\begin{table*}[tp]
 \centering
  \begin{tabular}{rrcc}
    \multicolumn{4}{c}{Table 1}\\
    \multicolumn{4}{c}{Properties of the Nebula around MR~2251$-$178}\\
    \tableline
    \tableline
    & & $log(U)$\tablenotemark{b} & $Z/Z_{\odot}$\tablenotemark{c}\\
    \tableline
    \multirow{2}{*}{Inner Cones\tablenotemark{a}} & Dusty & -1.0: [-2.0:, 0.0:] & 1.5: [1.0:, 2.0:]\\
    & Dust-Free & -2.5 [-3.0, -2.0] & 2.0 [1.0, 4.0]\\
    \tableline
    \multirow{2}{*}{Outer Cones\tablenotemark{a}} & Dusty & -1.0: [-2.0:, 0.0:] & 0.8: [0.5:, 1.5:]\\
    & Dust-Free & -2.2 [-3.0, -2.0] & 1.0 [0.5, 4.0]\\
    \tableline
    \multirow{2}{*}{Inner Anti-cones\tablenotemark{a}} & Dusty & -2.5 [-3.0, -2.0] & 1.5 [1.0, 2.0]\\
    & Dust-Free & -2.8 [-3.2, -2.5] & 0.8 [0.5, 1.2]\\
    \tableline
    \multirow{2}{*}{Outer Anti-cones\tablenotemark{a}} & Dusty & -2.7 [-3.1, -2.0] & 0.6 [0.3, 1.0]\\
    & Dust-Free & -3.2 [-3.5, -3.0] & 0.5 [0.2, 1.0]\\
    \tableline
  \end{tabular}
  \tablenotetext{a}{These regions are shown in Figure~3$a$. 
    The inner cones, outer cones, inner anti-cones, and outer anti-cones 
    correspond to the red, light blue, dark blue, and green crosses in 
    this figure, respectively.}
  \tablenotetext{b}{Ionization parameter defined in Equation~1. The typical uncertainty on these values is 
    $\pm$0.2 dex or $\pm$0.4 dex for entries followed by a colon (the dusty models do not fit  well the line ratios in the inner and outer cones).}
  \tablenotetext{c}{Metallicity relative to solar. The uncertainty on these values is $\pm$0.2 $Z_\odot$ or $\pm$0.4 $Z_\odot$  for entries followed by a colon. }
\end{table*}

The ionization level of the gas in the outer anti-cone (the green
crosses in Figure 3) drops considerably.  At these lower
[\ion{O}{3}]/H$\beta$ ratios, we must consider the possibility that
some of the ionizing radiation arises from star formation.  Using the
dilution curves of \citet{2010ApJ...709..884Y}, we estimate that
dilution by star formation is $\sim$50\% so the AGN contribution to
the ionizing radiation in this region may be underestimated by a
factor of up to two.

\subsection{Electron Densities, Ionizing Fluxes, and Escape Fractions}

The ionization parameters derived in \S 4.2 can be used to constrain
the fraction of the quasar ionizing flux that is escaping the nebula
and contributing to the ionization of the IGM. Given equation (1), we
first need to know the electron density in the nebula. For this we
examine discrete clumps in the nebula and use
\begin{equation}
L(H\alpha) = 2.85 \times n_en_p\alpha^{eff}_{H\beta}h\nu_{H\beta}V\epsilon, \label{eqn:eqn2}
\end{equation}
where $\alpha^{eff}_{H\beta}$ is the effective H$\beta$
recombination coefficient \citep[$3.03\times 10^{-14}$ cm$^3$ s$^{-1}$
for $T=10^4K$;][]{1989agna.book.....O}, $V$ is the volume of each
clump [assumed to be spherical with size corrected in quadrature for
the seeing ($\sim$0.7\arcsec\ FWHM)], and $\epsilon$ is the volume
filling factor (assumed to be unity in these clumps). For this
exercise, we use clumps that are approximately circular on the sky so
our assumption of spherical symmetry is reasonable. The
seeing-corrected clump diameters range from $\sim$1.0 to 5.0 kpc and
the densities are $n_e\sim 0.1-1.0~cm^{-3}$ (top panel of Figure~4),
showing a distinct enhancement within the inner $\sim$15 kpc. These
values for $n_e$ fall within the range derived by
\citet{1990ApJ...356..389M}. These densities are well below the
value assumed in the \citeauthor{2004ApJS..153....9G} models, but we
believe the ionization parameters derived from these models are
reliable (e.g., the critical density for collisional de-excitation is
8 $\times$ 10$^4$ and 7 $\times$ 10$^5$ cm$^{-3}$ for [\ion{N}{2}]
$\lambda$6583 and [\ion{O}{3}] $\lambda$5007, respectively).

Combining these electron densities with the ionization parameters
derived in \S 4.2, we calculate the flux of ionizing photons,
$\Phi_{measured}$ (photons s$^{-1}$ cm$^{-2}$), at various radii in
the bicone.  For comparison, we estimate a predicted ionizing photon
flux, $\Phi_{predicted}$, integrating from 1 to 10 Rydbergs the
power-law fit ($\alpha_\nu = -1.3$) to the far-UV nuclear spectral
energy of MR~2251$-$178 reported in \citet{2004ApJ...615..135S}. In
deriving the predicted flux, we do not account for UV absorption along
our line of sight, even though strong $Ly\alpha$ absorption is present
\citep{2001ApJ...559..675M}; $\Phi_{predicted}$ is thus likely a lower
limit.

The values of $\Phi_{predicted}$, $\Phi_{measured}$, and
$\Phi_{measured}/\Phi_{predicted}$ in the ionization cones are plotted
in Figure~4. We see no trend for this ratio to decrease with
increasing distance from the quasar. The ionizing radiation from the
quasar thus propagates through the nebula with negligible attenuation.
Since our deeper observations did not reveal line-emitting material
beyond what was already known by \citet{1999ApJ...524L..83S}, we
conclude that the nebula is matter-bounded {\em i.e.} all of the
ionizing radiation emitted along the ionization cones escapes the
nebula.

\begin{figure}[t]
  \epsscale{0.77}
  \plotone{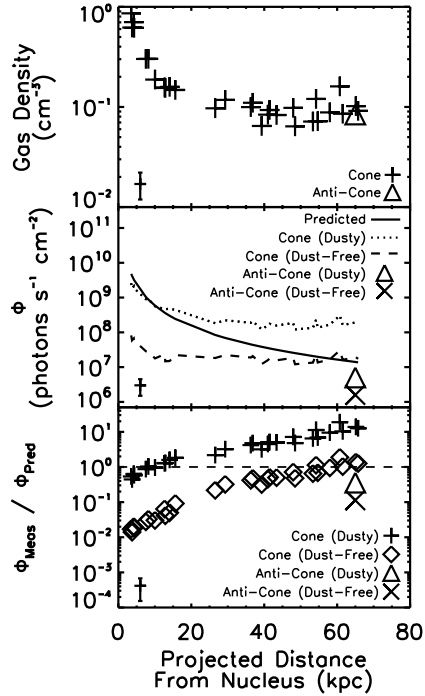}

  \caption{ The top panel shows the electron densities of discrete
    clumps within the ionization cones and anti-cones. The middle
    panel compares the predicted flux of ionizing photons, derived
    from the UV luminosity of the nucleus, with the measured flux of
    ionizing photons, derived from the clump electron densities and
    ionization parameters based on the dusty and dust-free models.
    The lower panel shows the ratios of measured to predicted ionizing
    photon fluxes for the dusty and dust-free cases.  The typical
    uncertainties are shown in the lower left corner of each panel.  }
  \label{fig:fig4}
\end{figure}

There is only one well-defined clump in the outer anti-cone region.
The inferred ionizing flux at that position is distinctly lower than
in the cone region (Figure 4). While this result may indicate that the
nebula is radiation-bounded in the anti-cone direction, dilution by
star formation may also be responsible for this apparent drop in
$\Phi_{measured}$, as discussed in \S 4.2.

\subsection{Implications}

Our results have important implications on the issue of AGN radiative
feedback.  First, we have shown that the ionizing radiation field of
MR~2251$-$178 is collimated along the east-west radio axis, producing
a bicone with opening angle $\sim$80$\degr -$130$\degr$ that covers
$\sim$65$-$90\% of 4$\pi$ steradians (these numbers take into account
the projection effects). Ionizing photons are leaking along directions
outside these cones at a rate that is $\sim$2-3 times lower than along
the cones. The nebula around MR~2251$-$178 appears to be
matter-bounded (except perhaps in the anti-cone direction, where we
have poor constraints; \S 4.3), indicating that the ionizing radiation
that makes it out of the quasar host galaxy also makes it out of the
nebula and contributes to ionizing the IGM.  A similar situation was
observed in HE~1029$-$1401, a luminous radio-quiet quasar with a
photoionized biconical nebula that extends out to 16 kpc
\citep{2010A&A...519A.115H}.  In MR~2251$-$178, however, the walls at
the base of the biconical structure are porous, allowing an additional
$\sim$5--15\% of the total ionizing flux from the quasar to escape
outside of the ionization cones. Overall, considering the
contributions from both inside and outside the cones, and the
possibility that the nebula is ionization-bounded outside the cones,
we find that $\sim$65--95\% of the quasar ionizing radiation makes its
way out of the system.

The situation may be different at higher redshift where galaxy hosts
have larger gas fractions on average than their local counterparts
\citep[e.g.,][]{2010Natur.463..781T, 2010ApJ...713..686D}. Dependences
on AGN luminosity likely also become important. Flux-limited AGN
surveys preferentially select luminous quasars which are more likely
to ionize the host ISM \citep[e.g.][]{2012ApJ...759..117C}, create
giant halos \citep[e.g.][]{2006MNRAS.370.1372F, 2007ApJ...657..135C,
  2011AJ....142..186W, 2012A&A...542A..91N, 2012MNRAS.425.1992C}, and
ionize the surrounding IGM out to Mpc scales
\citep[e.g.][]{2008ApJ...676..816G}.  The relatively small fraction of
type 2 quasars at high redshifts in current surveys suggests that the
opening angles of quasar ionization cones increase with increasing
quasar luminosities \citep[e.g.][]{1991MNRAS.252..586L}, raising the
efficiency of quasars to ionize the IGM. A similar effect may be
taking place in high-redshift star-forming galaxies where giant
Ly$\alpha$ halos have been detected \citep{2011ApJ...736..160S} and
galactic winds poke holes through the ISM where Lyman continuum
emission can escape \citep[][and references
therein]{2013ApJ...765...47N} \citep[c.f. nearby dwarf
galaxies:][]{2011ApJ...741L..17Z}.  A recent study by
\citet{2013MNRAS.430.2327L} finds multiple cases of smoothly
distributed ionized envelopes around luminous radio-quiet quasars at
$z\sim0.5$.  Line ratio maps similar to those presented in this paper
would help refine our understanding of the geometry and escape
fraction of the ionizing radiation fields in these objects,
hence the role these luminous quasars play in ionizing the IGM.

\acknowledgements

We thank M. McDonald for help with data acquisition and reduction, and
the referee, M. Villar-Mart\'{\i}n, for a thoughtful report.  This
work was funded through NSF grant AST-10009583 (K.K., S.V.),
a Senior NASA Postdoctoral Program award (S.V.), and the Alexander von
Humboldt Foundation (S.V.).  

\end{document}